\documentclass[12pt]{article}
\usepackage{indentfirst,psfig}
\usepackage{doublespace}

\begin{document}

\title{Strategies for High-Throughput, Templated Zeolite Synthesis}
\author{
Ligang Chen and Michael W. Deem\\
Department of Chemical Engineering\\
University of California, Los Angeles, CA  90095--1592
}

\maketitle

\centerline{\textbf{Abstract}}
\begin{quote}
How best to design and redesign high-throughput experiments for
zeolite synthesis is addressed.  A model that relates materials
function to chemical composition of the zeolite and the structure
directing agent is introduced.  Using this model, several Monte
Carlo-like design protocols are evaluated.  Multi-round protocols are
found to be effective, and strategies that use \emph{a priori}
information about the structure-directing libraries are found to be
the best.
\end{quote}

\newpage

\section{Introduction}
Zeolites have found wide industrial applications in processes such as
catalysis, molecular sieving, gas separation, and ion exchange. The significant
catalytic activity and selectivity of zeolite materials are attributed
to the large internal surface area and highly distributed active sites
that are accessible through uniformly-sized pores. However, due to the
desire to improve the quality of known zeolites that are small or
faulted and the continuing quest for new zeolites with different
functionalities, novel syntheses of zeolites with various frameworks and
chemical compositions have drawn heavy research attention. To date,
roughly 150 framework structures have been reported. New structures
are added at the pace of roughly 6 per year. The use of organo-cation
template molecules to provide structure direction has given rise to
breakthroughs in zeolite science and provides much of the impetus for
current synthetic efforts~\cite{Zones98,Davis97}. The addition of
organic templates such as amines and alkylammonium ions to zeolite
synthesis gels can affect the rate at which a particular material is
formed and can make new structures or framework chemical compositions
accessible. In some cases, a strong correlation between the geometry
of the structure directing organic molecule and the zeolite pore
architecture is found. The nature and extent of interactions between
the organic templates and the inorganic components of a zeolite synthesis
gel are key parameters that determine the final zeolite pore
architecture~\cite{Davis951,davis952}.

We are interested in high-throughput synthesis of
zeolites. High-throughput, or combinatorial, methods allow for
simultaneous creation of a large number of structurally diverse and
complex compounds, generalizing the traditional techniques of single
compound synthesis. Monte Carlo methods have been proposed and shown
to be efficient methods for library design and re-design in both
material discovery~\cite{Falcioni00,Deem01} and small molecule
design~\cite{Chen01}. Material discovery deals with continuous
variables, such as composition variables and non-composition
variables. Small molecule design deals with discrete variables, such
as the identities of template and ligand. For templated zeolite
synthesis, we have both the continuous zeolite composition and
non-composition variables and the discrete identity variables of the
component parts of the organic structure-directing molecules. All of
these variables affect the final zeolite material in an interrelated
way.

We here propose several strategies for templated, high-throughput
zeolite synthesis. In section 2 we introduce a new random energy model
that serves as a surrogate for experimental measurement of the figure
of merit.  This random energy model is based upon models used in
material discovery and small molecule design.  The random energy model
is not fundamental to the protocols; it is introduced as a simple way
to test, parameterize, and validate the various searching methods. In
an experimental implementation, the random energy model will be
replaced by the value returned by the screen. In section 3, we
introduce several protocols for experimentation. In section 4, we
compare results of each protocol and discuss some implications.
Finally, we conclude in section 5.

\section{Random Energy Model}
To test proposed protocols in an efficient fashion, we need a model in
lieu of the real experimental screening process.  That is, we need a
model that relates the chemical composition of the zeolite and
structure directing agent to the function, or figure of merit, of the
material.  This model is not essential to the protocols. It is simply
a fast and cheap means to test proposed protocols. This model will
capture the essence of the physical system and provide validation for
the protocols. Random energy models have
proven useful for the study of several complex fields, including protein
folding~\cite{Bryngelson,Shakhnovich}, protein molecular
evolution~\cite{Deem99}, material discovery~\cite{Falcioni00}, and
small molecule design~\cite{Chen01}.

To consider template-assisted zeolite synthesis, we use a model that
combines ideas from both material discovery~\cite{Falcioni00} and
small molecule design~\cite{Chen01}. The value of the figure of merit
is naturally a sum of a zeolite energy and a well-chosen organic
molecular energy:
\begin{equation}
E = E_{\rm zeolite} + E_{\rm molecule}
\end{equation}

\begin{figure}[htbp]
\centering
\leavevmode
\psfig{file=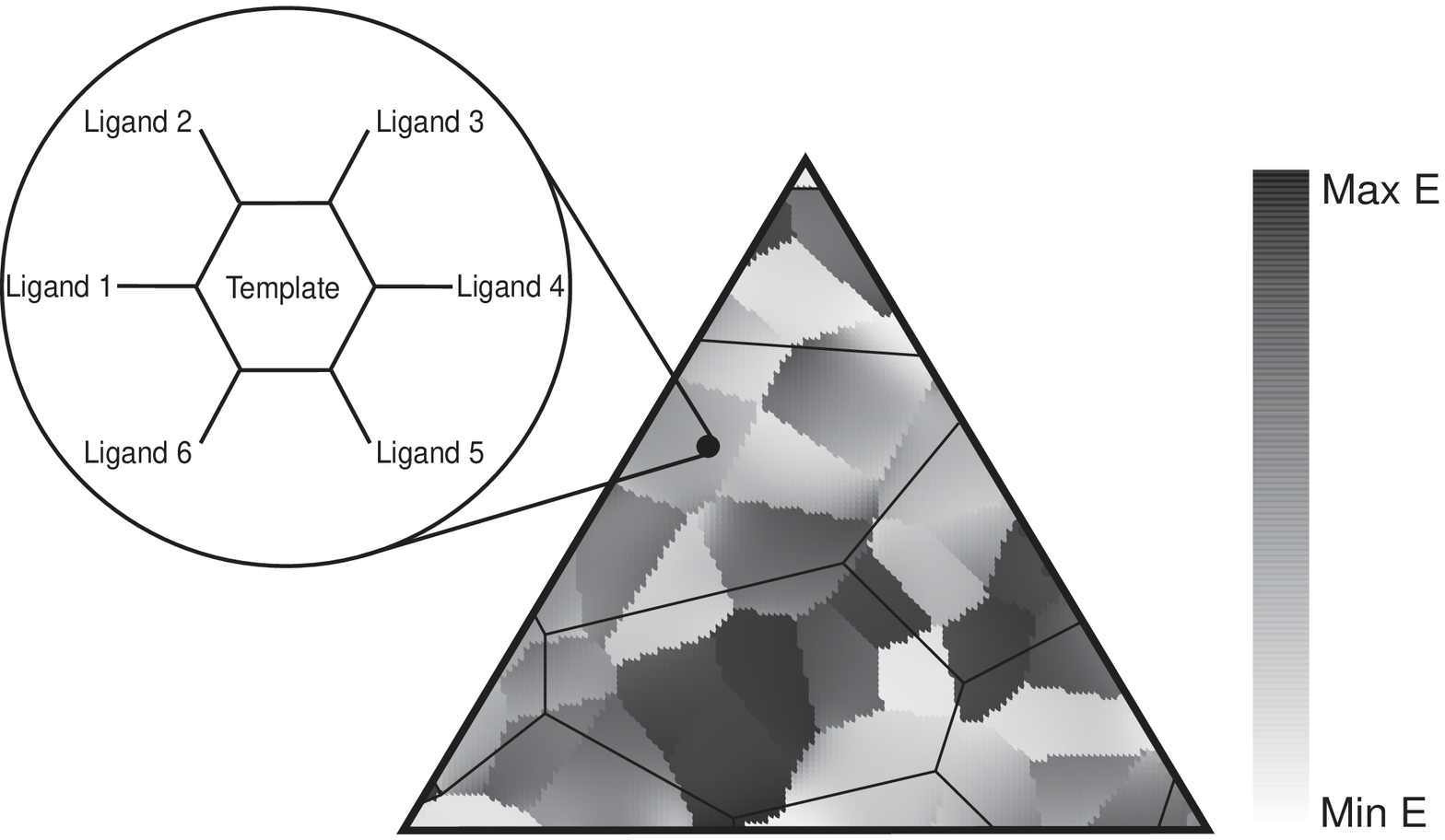,height=4in}
\caption{Schematic of the random energy model for templated zeolite synthesis.}
\label{fig1}
\end{figure}

The zeolite framework composition variables $x_i$, $d$ in number, are
certainly key variables in $E_{\rm zeolite}$. The composition
variables are specified by the mole fractions $x_i$, with $0 \le x_i
\le 1 $, and $\sum_{i=1}^{d} x_i = 1$.  Typical elemental compositions
in zeolite frameworks include silicon, oxygen, aluminum, phosphorus,
boron, and germanium, among others.  As discussed in the Appendix, the
allowed compositions are constrained to lie within a simplex in $d-1$
dimensions. Several phases will exist for different compositions of
the material. The figure of merit will be dramatically different
between each of these distinct phases. To mimic this behavior, the
composition variables are grouped in the Random Phase Volume Model
into phases centered around $N_x$ points ${\bf x}_{\alpha}$ randomly
placed within the allowed composition range. The phases form a Voronoi
diagram, see Figure~\ref{fig1}. The model is defined for any number of
composition variables, and the number of phase points is defined by
requiring the average spacing between phase points to be $\xi = 0.25$.
To avoid edge effects, additional points are added in a belt of width
$2 \xi$ around the simplex of allowed compositions.  The figure of
merit should change dramatically between composition phases.
Moreover, within each phase $\alpha$, the figure of merit should also
vary with ${\bf x} - {\bf x}_\alpha$ due to crystallinity effects such
as crystallite size, intergrowths, defects, and
faulting. Non-composition variables should also affect the measured
figure of merit. The non-composition variables are denoted by the $b$
variables $z_i$. Without loss of generality, each variable ranges $-1
\leq z_i \leq 1$. How the figure of merit changes with the
non-composition variables should depend on the value of the
composition variables. The non-composition variables fall within $N_z$
non-composition phases. There are a factor of 10 fewer non-composition
phases than composition phases.  When ${\bf x}$ is in composition
phase $\alpha$ and non-composition phase $\gamma$, the zeolite energy
is given by
\begin{equation}
E_{\rm zeolite}({\bf x}, {\bf z}) = H({\bf x}-{\bf x}^{\alpha},
\{G^{\alpha}\}) + \frac{1}{2} H({\bf z}, \{G^{\gamma}\})
\end{equation}
where $\{G^{\alpha}\}$ and $\{G^{\gamma}\}$ are phase-dependent random
Gaussian variables with zero mean and unit variance. The
$\{G^{\alpha}\}$ are different in each composition phase, and the
$\{G^{\gamma}\}$ are different in each non-composition phase. The form
of the function $H$ is
\begin{equation}
H(w_1, \ldots, w_n, \{G\}) = G_0 + \sigma_H
\sum_{k=1}^{q}~\sum_{\stackrel{i_1+ \ldots +i_n=k}{i_1, \ldots,
i_n\geq 0}} f_{i_1 \ldots i_n; k} \xi_H^{-k}G_{i_1 \ldots
i_n}w_1^{i_1} \ldots w_n^{i_n} \label{eqn:H}
\end{equation}
The degree of the polynomial is chosen as $q = 6$.
The set of coefficients $\{G\}$ includes
$G_0$ and $G_{i_1 \ldots i_n}$.
The constant symmetry factors
$f_{i_1 \ldots i_n; k}$ are given by
\begin{equation}
f_{i_1 \ldots i_n; k} = \frac{k!}{i_1! \ldots i_n!}
\end{equation}
The scale factor $\xi_H$ is chosen so that each term in the
multinomial contributes roughly the same amount
in the root-mean-square sense.
 For the composition
variables $\xi_H = \xi / 2$, and for the non-composition variables
$\xi_H = (\langle z^6 \rangle / \langle z^2 \rangle)^{1/4}
=(3/7)^{1/4}$. The $\sigma_H$ are chosen so that the multinomial terms
contribute $40\%$ as much as the corresponding constant, phase terms,
$G_0$ in eq~\ref{eqn:H}, 
in the root-mean-square sense.

The organic structure-directing molecules are uniquely
characterized by composition, such as the identity of the
template and ligands, as shown in Figure \ref{fig1}.
Note that \emph{template} here denotes the framework of the
structure-directing molecule, onto which ligands are attached, not
the entire structure-directing molecule.
 For simplicity, we consider the small molecule
to consist of one template and six binding ligands, each drawn from
a fragment library. To quantitatively describe each template
and ligand, six weakly correlated descriptors are used. Their numeric
values specify the characteristics of each template or ligand.

We again use a random energy model for the molecules to account for
the contributions to structure direction arising from interactions between
the zeolite and template and from interactions between the zeolite
and each of the ligands. In addition, synergistic effects between the
ligands and template are incorporated. Consider, for example, a
molecule made from template number $m$ from the template library and
ligands number $s_1, \ldots, s_6$ from the ligand library. The
template is characterized by six descriptors, $D_1^{(m)}, \ldots,
D_6^{(m)}$. Similarly, each ligand is characterized by six
descriptors, $d_1^{(l_i)}, \ldots, d_6^{(l_i)}$. We denote the
template contribution to binding by $E_{\rm T}$ and the ligand
contributions by $E_{\rm L}$.  We denote the contribution due to
synergistic ligand-ligand interactions by $E_{\rm LL}$ and the
contribution due to synergistic template-ligand interactions by
$E_{\rm TL}$. The total molecular contribution to the figure of merit
is, then,

\begin{equation}
E_{\rm molecule} = E_{\rm L}+E_{\rm T}+E_{\rm LL} + E_{\rm TL}
\end{equation}
Each of these terms is given in the form of a random polynomial:
\begin{eqnarray}
E_{\rm L}({\bf x}, {\bf z}) &=& \sum_{i=1}^6 F[d_1^{(l_i)}, \ldots,
d_6^{(l_i)}, \{G_{\rm L}({\bf x}, {\bf z})\}],~~~G_{\rm L}({\bf x},
{\bf z}) = \lambda_{x, 1} G_{\rm L}^{\alpha} + 
\frac{1}{2} \lambda_{z, 1} G_{\rm L}^{\gamma}
\label{eqn:l}\\
E_{\rm T}({\bf x}, {\bf z}) &=& F[D_1^{(m)},\ldots, D_6^{(m)},
\{G_{\rm T}({\bf x}, {\bf z})\}],~~~G_{\rm T}({\bf x}, {\bf z}) =
\lambda_{x, 2} G_{\rm T}^{\alpha} + \frac{1}{2} 
\lambda_{z, 2} G_{\rm T}^{\gamma}\\ 
E_{\rm LL}({\bf x}, {\bf z}) &=& \sum_{i=1}^{6} h_i({\bf x}, {\bf z}) F
[d_{j_1}^{(l_i)}, d_{j_2}^{(l_i)}, d_{j_3}^{(l_i)}, d_{j_4}^{(l_{i+1})},
d_{j_5}^{(l_{i+1})}, d_{j_6}^{(l_{i+1})}, \{G_{\rm LL}({\bf x}, {\bf
z})\}], \nonumber \\ 
& & G_{\rm LL}({\bf x}, {\bf z}) = \lambda_{x, 3} G_{\rm LL}^{i,
\alpha} + \frac{1}{2} \lambda_{z, 3} G_{\rm LL}^{i, \gamma}
\label{eqn:ll}\\
E_{\rm TL}({\bf x}, {\bf z}) &=& \sum_{i=1}^{6} h_i({\bf x}, {\bf z})
F[d_{k_1}^{(l_i)}, d_{k_2}^{(l_i)}, d_{k_3}^{(l_i)}, D_{k_4}^{(m)},
D_{k_5}^{(m)}, D_{k_6}^{(m)}, \{G_{\rm TL}({\bf x}, {\bf z})\}],\nonumber \\
& &G_{\rm TL}({\bf x}, {\bf z}) = \lambda_{x, 4} G_{\rm TL}^{i, \alpha}
+ \frac{1}{2} \lambda_{z, 4} G_{\rm TL}^{i, \gamma}
\label{eqn:tl}
\end{eqnarray}
where individual descriptors from each template and ligand are given
in corresponding libraries. The building block for our random energy
model as a function of those descriptors is
\begin{equation}
F(w_1, \ldots, w_n, \{G\}) = \sum_{k=0}^{q}~\sum_{\stackrel{i_1+ \ldots
+i_n=k}{i_1, \ldots, i_n\geq 0}} f_{i_1 \ldots i_n; k}^{\frac{1}{2}}
\xi_F^{-k}G_{i_1 \ldots i_n}w_1^{i_1} \ldots w_n^{i_n}
\end{equation}
The $\{G\}$, again, will be composed of a composition and a
non-composition piece, as indicated in
eqs~\ref{eqn:l}--\ref{eqn:tl}. We use $q=6$ and $n = 6$. Since the
values of the descriptors, $w_i$, will be drawn from a Gaussian random
distribution of zero mean and unit variance, we set the scaling
factor, $\xi_F$, by
\begin{equation}
\xi_F=\left(\frac{\langle w^q \rangle} {\langle w^2
\rangle} \right)^{\frac{1}{q-2}} = \left(\frac{q!}{(q/2)!2^{q/2}}
\right)^{\frac{1}{q-2}}
\end{equation}
In general the polynomial coefficients, $\{ G \}$ in eq
(\ref{eqn:l}-\ref{eqn:tl}), are functions of both zeolite variables
${\bf x}$ and ${\bf z}$. However, for simplicity, we assume that the
coefficients are
phase-dependent only. Both types of zeolite phases have a significant
impact on the energy function. The variables $\{G_{\rm L}^{\alpha}\},
\{G_{\rm L}^{\gamma}\}, \{G_{\rm T}^{\alpha}\}$, $\{G_{\rm
T}^{\gamma}\}, \{G_{\rm LL}^{i, \alpha}\}, \{G_{\rm LL}^{i, \gamma}\},
\{G_{\rm TL}^{i, \alpha}\}$, and $\{G_{\rm TL}^{i, \gamma}\}$ are
fixed random Gaussian variables with zero mean and unit variance.
These coefficients are different in each phase. The structural
constant $h_i$ indicates the strength of the interaction at structure
directing site $i$. This interaction function depends on
the  zeolite variables ${\bf x}$ and
${\bf z}$ as
\begin{equation}
h_i({\bf x}, {\bf z}) = H({\bf x}-{\bf x}^{\alpha}, \{G^{\alpha}_i\}) +
\frac{1}{2} H({\bf z}, \{G^{\gamma}_i\})
\end{equation}
As before $\{G^{\alpha}_i\}$ and $\{G^{\gamma}_i\}$ are sets of
phase-dependent fixed random Gaussian variables with zero mean and
unit variance. The parameter $\sigma_H$ is adjusted so that the
multinomial terms contribute roughly the same as the constant, phase
terms on average. In $E_{\rm LL}$, only synergistic interactions
between neighboring ligands are considered, and it is understood that
$l_7$ refers to $l_1$ in eq~\ref{eqn:ll}. In principle, the polynomial
in eq~\ref{eqn:ll} could be a function of all 12 descriptors of both
ligands. We assume, however, that important contributions come from
interactions among three randomly chosen distinct descriptors of
ligand $l_i, d_{j_1}^{(l_i)}, d_{j_2}^{(l_i)},$ and $d_{j_3}^{(l_i)},$
and another three randomly chosen distinct descriptors of ligand
$l_{i+1}, d_{j_4}^{(l_{i+1})}, d_{j_5}^{(l_{i+1})},$ and
$d_{j_6}^{(l_{i+1})}$. Similarly, we assume that template-ligand
contributions come from interactions between three randomly chosen
distinct descriptors of the ligand, $d_{k_1}^{(l_i)},
d_{k_2}^{(l_i)},$ and $d_{k_3}^{(l_i)},$ and another three randomly
chosen distinct descriptors of the template, $D_{k_4}^{(m)},
D_{k_5}^{(m)},$ and $D_{k_6}^{(m)}$. Both $j_i$ and $k_i$ are
descriptor indices ranging from 1 to 6.  We assume that these indices
depend only on the template, since the molecular details of the
zeolite have been suppressed.  The constants $\lambda_{x, i}$ are
adjusted so that the synergistic terms in compositional phases will
contribute on average as $E_{\rm L}:E_{\rm T}:E_{\rm LL}:E_{\rm TL} =
1:1:2:1.2$. Furthermore, we adjust the $\lambda_{x, i}$ so that the
total molecular contribution from the interaction with the
compositional variables accounts for roughly 15\% of the constant,
compositional phase term in $E_{\rm zeolite}$ in the root-mean-square
sense. We similarly adjust the values of $\lambda_{z, i}$ in
non-compositional phases so that the contribution from the
interaction with the non-composition variables is also 15\%.

Finally, we minimize the energy returned by the model.  That is,
we sample the figure of merit, $E$, by $\exp{(-\beta E)}$.

\section{Protocols for high-throughput experimentation}
As in the small molecule case~\cite{Chen01}, we first build the
template and ligand libraries. We denote the size of the template
library by $N_{\rm T} = 15$ and the size of the ligand library by
$N_{\rm L} = 1000$. In a real experiment, the six descriptors would
then be calculated for each template and ligand. In the simulated
experiment, the values of the six descriptors of each ligand and
template are extracted from a Gaussian random distribution with zero
mean and unit variance. In the simulated experiment, we also associate
two sets of random interaction descriptor indices to each template for
the interaction terms in eq~\ref{eqn:ll} and~\ref{eqn:tl}.

For comparison, we fix the total number of samples to be synthesized
at 100000 for all protocols. That is, we keep roughly the same
experimental cost.

We consider several single pass protocols.  Current high-throughput
experiments uniformly tend to perform a single-pass, grid search on
all continuous variables. To mimic current experiments, we use such a
grid search on the composition and non-composition variables. For the
discrete molecular variables, we instead randomly pick one template
and six ligands from the corresponding libraries. In another
single-pass protocol, which we call random, we search randomly over
all variables, \emph{i.e.}\ we choose the composition variables and
non-composition variables at random as well.

We also consider several multi-pass protocols.
Unlike single-pass protocols, multi-pass protocols allow us to
learn about the system as the experiment proceeds. In analogy to Monte
Carlo methods, each round of combinatorial chemistry experiments
corresponds to a move in a Monte Carlo simulation. Instead of tracking
one system with many configurational degrees of freedom, however, many
samples are tracked. This results in a rather diverse population of
samples and increases the opportunity for a few 
zeolite samples to survive
more elaborate tests for application performance, tests that
are only roughly correlated with the primary screen.
The existence of these secondary and primary screens is the reason
why sampling the figure of merit to find several promising compounds
is so important.
We use a Monte Carlo protocol so as to
maintain diversity in the samples. Genetic algorithms, while 
powerful at local optimization, reduce the diversity of the
population in the cross-over step and tend to fail diversity
tests~\cite{Chen01}. To be consistent with current experiments, we
prefer to synthesize 1000 samples simultaneously and perform Monte
Carlo moves on every sample for 100 rounds. The initial sample
configurations are assigned by the random protocol. 

We change the zeolite variables as well as the components of the
structure directing agent at each round of the Monte Carlo protocol.
In the simplest approach, we perturb both the composition variables
and the non-composition variables of the zeolite part around the
original value using the traditional Metropolis-type method. For the
non-composition variables, ${\bf z}$, periodic boundary conditions are
used. For the composition variables, ${\bf x}$, reflecting boundary
conditions are used, as discussed in the Appendix. For the molecular
part, either the template is changed with probability $p_{\rm
template}$, or one of the six ligands is picked randomly to change in
each Monte Carlo move. The new configurations are updated according to
the acceptance rule at $\beta$, the inverse of the protocol
temperature. All the samples are sequentially updated in one Monte
Carlo round. For the molecular part, either a simple Metropolis method
or a biased Monte Carlo method can be incorporated. In the simple
Metropolis method, the new template or ligand is selected at random
with a uniform probability from the corresponding libraries. The final
acceptance rule for an unbiased protocol,
 which is applied after the zeolite and molecular
variables have been modified, is
\begin{equation}
{\rm acc}({\rm o} \rightarrow {\rm n})=\min[1,~\exp(-\beta \Delta E)]
\end{equation}

The biased Monte Carlo schemes allow us to generate a trial configuration
with a probability that depends on the potential energy of that
configuration. These schemes have been proven to be highly successful
in small molecule high-throughput experiment design~\cite{Chen01},
where biased energy forms are constructed from either theory or
pre-experiments. In the present case, since the total molecular energy
is a function of both the composition and non-composition variables,
it is not feasible to construct the bias from pre-experiments. We,
therefore, use the theoretical bias directly from the random energy model.
 For a
zeolite at compositional phase $\alpha$ and non-compositional phase
$\gamma$, the bias energy form for template $m$ is
\begin{equation}
E^{(m)}({\bf x}, {\bf z}) = F[D_1^{(m)},\ldots, D_6^{(m)},
\{G_{\rm T}({\bf x}, {\bf z})\}],~~~~~
      G_{\rm T}({\bf x}, {\bf z})= 
     \lambda_{x, 2} G_{\rm T}^{\alpha} + \frac{1}{2} \lambda_{z, 2}
G_{\rm T}^{\gamma}
\end{equation}
And the bias for ligand $i$ is
\begin{equation}
e^{(i)}({\bf x}, {\bf z}) = F[d_1^{(i)}, \ldots, d_6^{(i)}, 
      \{G_{\rm  L}({\bf x}, {\bf z}) \}],~~~~~
    G_{\rm  L}({\bf x}, {\bf z}) = \lambda_{x, 1} G_{\rm
L}^{\alpha} + \frac{1}{2} \lambda_{z, 1} G_{\rm L}^{\gamma}
\end{equation}
These biases tend to exhibit a large gap between a few dominant
templates and ligands and the rest. To ensure the participation of
more ligands and templates in the strategy, we introduce cutoff
energies for the ligand and template, $e_{\rm c}$ and $E_{\rm c}$.  We
choose $e_{\rm c}$ to be the 21st lowest ligand energy and
$E_{\rm c}$ to be the 4th lowest template energy. With this cutoff
energy, the biased energy, $e_{\rm b}^{(i)}$, for the $i^{th}$ ligand
becomes
\begin{equation}
e_{\rm b}^{(i)}({\bf x}, {\bf z})=\left\{ \begin{array}{ll}
                e^{(i)}({\bf x}, {\bf z}) & \mbox{if $e^{(i)}>e_{\rm c}$} \\
                e_{\rm c}({\bf x}, {\bf z}) & \mbox{otherwise}
                \end{array}
        \right.
\end{equation}
And the biased energy, $E_{\rm b}^{(m)}$, for the $m^{th}$ template
become
\begin{equation}
E_{\rm b}^{(m)}({\bf x}, {\bf z})=\left\{ \begin{array}{ll}
                E^{(m)}({\bf x}, {\bf z}) & \mbox{if $E^{(m)}>E_{\rm c}$} \\
                E_{\rm c}({\bf x}, {\bf z}) & \mbox{otherwise}
                \end{array}
        \right.
\end{equation}

In one Monte Carlo move, if the $i^{th}$ ligand in the molecule is to be
changed at new zeolite values ${\bf x}'$ and ${\bf z}'$, the biased
probability for selecting another ligand $l_i'$ from the ligand
library is
\begin{equation}
f[E({\rm n})] = \frac{\exp[-\beta e_{\rm b}^{(l_i')} ({\bf x}', {\bf z}')]}
{\sum_{j=1}^{N_{\rm L}} \exp[-\beta e_{\rm b}^{(j)}({\bf x}', {\bf
z}')]}
\end{equation}
For reverse move, we have
\begin{equation}
f[E({\rm o})] = \frac{\exp[-\beta e_{\rm b}^{(l_i)} ({\bf x}, {\bf z})]}
{\sum_{j=1}^{N_{\rm L}} \exp[-\beta e_{\rm b}^{(j)}({\bf x}, {\bf
z})]}
\end{equation}

Similarly, if the template is desired to be changed at new zeolite
values ${\bf x}'$ and ${\bf z}'$, the biased probability for selecting
template $m'$ is
\begin{equation}
f[E({\rm n})] = \frac{\exp[-\beta E_{\rm b}^{(m')} ({\bf x}', {\bf z}')]}
{\sum_{j=1}^{N_{\rm T}} \exp[-\beta E_{\rm b}^{(j)}({\bf x}', {\bf
z}')]}
\end{equation}
For reverse move, we have
\begin{equation}
f[E({\rm o})] = \frac{\exp[-\beta E_{\rm b}^{(m)} ({\bf x}, {\bf z})]}
{\sum_{j=1}^{N_{\rm T}} \exp[-\beta E_{\rm b}^{(j)}({\bf x}, {\bf
z})]}
\end{equation}

To satisfy the detail balance, the acceptance rule becomes
\begin{equation}
{\rm acc}({\rm o} \rightarrow {\rm n})
    = \min \left\{1, \frac{f[E({\rm o})]}{f[E({\rm n})]}
\exp(-\beta \Delta E) \right \}
\end{equation}

Parallel tempering is known to be a powerful tool for searching rugged
energy landscapes~\cite{Swendsen,Geyer,Falcioni99}. Parallel tempering can be
combined with both Metropolis Monte Carlo and biased Monte Carlo. In
parallel tempering, the samples are divided into $k$ groups. The first
group of samples is simulated at $\beta_1$, the second group is at
$\beta_2$, and so on, with $\beta_1< \beta_2 <\ldots < \beta_k$. At
the end of each round, samples in group $i<k$ are allowed to exchange
configurations with samples in group $i+1$ with probability $p_{\rm
ex}$. The acceptance rule for a parallel tempering
exchange move is
\begin{equation}
{\rm acc}({\rm o} \rightarrow {\rm n})=\min[1,~\exp(\Delta \beta~\Delta E)]
\end{equation}
where $\Delta\beta=\beta_i-\beta_{i+1}$ and $\Delta E$ is 
the difference in energy between the sample in group $i$ and the sample in
group $i+1$ before the exchange is made.
 It is important to notice that this exchange step is
experimentally cost-free. Nonetheless, this step can be dramatically
effective at facilitating the protocol to escape from local
minimum. The number of groups, the number of samples in each group,
the value of $\beta_i$, and the exchange probability, $p_{\rm ex}$, are
experimental parameters to be tuned.

\section{Results and Discussion}
The size of the library is fixed at $N_{\rm T} = 15$ and $N_{\rm L} =
1000$. We adjust $\lambda_{x, i}$ and $\lambda_{z, i}$ so that the
synergistic terms will contribute on average as $E_{\rm L}:E_{\rm
T}:E_{\rm LL}:E_{\rm TL} = 1:1:2:1.2$ in the composition and
non-composition phases. We also want a model in which the contribution
from zeolite and molecule will be roughly of the same order for the
optimal configurations obtained by grid or random protocols.  Due to
the complicated molecular terms, we find the parameters so that this
occurs by trial and error.  We first adjust all the $\sigma_H$ values,
then we adjust the $\lambda_{x, i}$ and $\lambda_{z, i}$ in
eqs~\ref{eqn:l}--\ref{eqn:tl}. As mentioned, we adjust the
$\lambda_{x, i}$ and $\lambda_{z, i}$ so that the total molecular
contribution from either type of phase accounts for roughly 15\%
of the corresponding constant phase term in $E_{\rm zeolite}$.

To locate optimal parameters for the protocols, we perform a few short
pre-experiments. Since the size of the template library is relatively
small, the optimal value for the probability of changing a template is
$p_{\rm template} = 0.02$. The maximum random displacements are
$|\Delta {\bf x}| = 0.1/\sqrt{d-1}$ and $|\Delta {\bf z}| = 0.2$ in
the composition space and non-composition space, respectively. For simple
Metropolis Monte Carlo, the optimal inverse protocol temperature is
$\beta = 50$ for X3 and X4 and $\beta = 20$ for X5. For the biased
Monte Carlo schemes, the optimal inverse protocol temperature is
$\beta_b = 500$ for X3 and X4 and $\beta_b = 200$ for X5. For biased
Monte Carlo combined with parallel tempering, it is optimal to have
the samples divided into three subsets, with 25\% of the population at
$\beta_1 = \frac{1}{2} \beta_b$, 50\% at $\beta_2 = \beta_b$, and 25\%
at $\beta_3 = 2 \beta_b$. The switching probability, $p_{\rm ex}$, is
0.1. Determination of these parameter values corresponds
experimentally to gaining familiarity with the protocol on a new
system.

\begin{figure}[htbp]
\centering
\leavevmode
\psfig{file=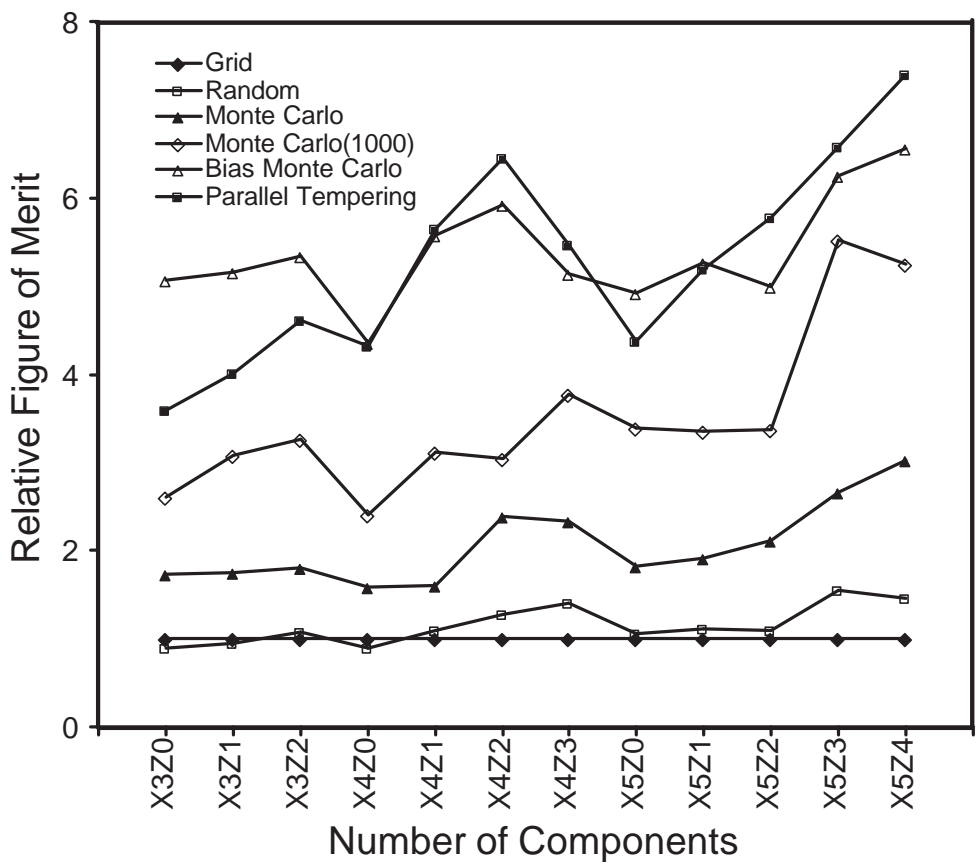,height=4.5in}
\caption{The minimum figure of merit found by different protocols
with different number of composition ({\bf x}) and non-composition
({\bf z}) variables. The complexity of the structure-directing agent
libraries is the same in all cases.  The results are scaled to the
minimum found by the grid searching method. Each value is averaged
over scaled results on 10 different instances.}
\label{fig2}
\end{figure}

All the protocols are tested with increasing numbers of composition
and non-composition variables. Results are shown in 
fig~\ref{fig2}. From this figure,
one can see that all multi-round Monte Carlo protocols are better than
the single-pass protocols such as grid and random. The simple Metropolis
method can find optimum samples that are twice as superior as those 
from the grid or random protocols. Most importantly, the biased Monte Carlo
is especially efficient. The optimum sample energies from biased Monte
Carlo far exceed the optimum sample energies from simple
Metropolis with even 1000 rounds. 
Biased Monte Carlo is able to find such favorable samples
mainly by finding
much more favorable designs for the structure-directing agent.
Parallel tempering is noticeably effective only
for the more complicated X4 and X5 systems. For the relatively simple systems,
such as X3, it is more important
to keep all 1000 of the samples at 
the optimum temperature.

\section{Conclusion}
We have presented a model for high-throughput zeolite synthesis. As in
previous studies, multi-pass Monte Carlo methods work better than do
single-pass protocols.  Sophisticated biased Monte Carlo schemes are
highly efficient and much better than simple Metropolis Monte Carlo.
Parallel tempering is the best method for complicated systems with 5
or more framework chemical compositional variables.

\section*{Acknowledgment}
It is a pleasure to acknowledge stimulating discussions with Daniel L.\ Cox
that lead this work.
This research was supported by the Alfred P.\ Sloan Foundation through
a fellowship to M.W.D., the Chevron Research and Technology
Corporation, and the National Science Foundation.

\section*{Appendix: reflecting boundary conditions}
The valid region of composition variables form a simplex, since they
are non-negative and sum to unity. For those
points near the corner of the allowed simplex, 
a high percentage of proposed
moves will go beyond the allowed region. If those moves are
simply rejected, the composition point
will get stuck in the corner for a long time, and the limited
sampling chances will be wasted.
 We want an algorithm that satisfies detail balance while avoiding
getting stuck in the corners.

\begin{figure}[htbp]
\centering
\leavevmode
\psfig{file=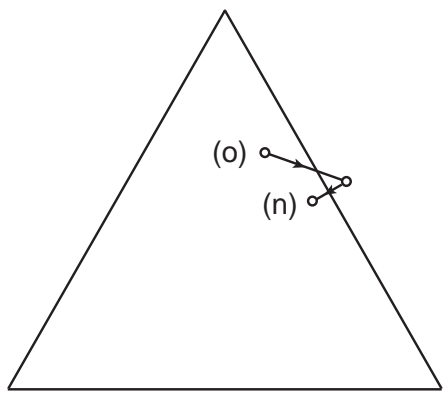,height=2in}
\caption{Schematic of the reflecting boundary conditions used to
obtain valid composition variables.}
\label{fig3}
\end{figure}

The scheme we have chosen ensures that all moves from any point in the
simplex lead to new valid points in the simplex.  The scheme basically
involves many reflections over the $d-2$ dimensional hyper-planes that
define the boundaries of the simplex.  Figure~\ref{fig3} shows how
this works in the case of $d=3$ for a single reflection.  We first
define the unit normals, ${\bf n}_i$, to each of the faces of the
simplex.  We also find the constants $c_i$ that define the faces by
the equation ${\bf x} \cdot {\bf n}_i = c_i$.  In the case of figure
\ref{fig3}, there are three such unit normals and three such
constants.  We initially define ${\bf y} = {\bf x}({\rm o})$.
The algorithm consists of the following steps:
\begin{enumerate}
\item Test whether the proposed new point, ${\bf x}$, is in the allowed
      region. If so, define the new point ${\bf x}({\rm n})$ to be ${\bf x}$
      and stop.
\item If not, find the face $i$ for which the quantity 
     $t_i = (c_i - {\bf y} \cdot {\bf n}_i) / 
          [ ({\bf x} - {\bf y}) \cdot {\bf n}_i]$
      is minimal, taking into account only those faces for which
      both $t_i$  and $({\bf x} - {\bf y}) \cdot {\bf n}_i$ are
      positive.
\item Define ${\bf y} \leftarrow {\bf y} + t_i ({\bf x} - {\bf y})$.
\item Reflect the point through this face by the equation ${\bf x} 
\leftarrow
     {\bf x} - 2 ( {\bf x} \cdot {\bf n}_i - c_i) {\bf n}_i$.
\item Continue with step 1.
\end{enumerate}
This algorithm converges, because it decreases the magnitude of ${\bf x}$ at
each step, until ${\bf x}$ is within the allowed region.
The algorithm is also reversible.  That is, for each forward move
${\bf x}({\rm o}) \to {\bf x}$ that maps to the new position
${\bf x}({\rm n})$, there is always a reverse move
${\bf x}({\rm n}) \to {\bf x}'$ that maps to the 
original position ${\bf x}({\rm o})$.  Furthermore, when the
original move is in the allowed move sphere,
$\vert {\bf x}({\rm o}) - {\bf x} \vert < \Delta x$, the
reverse move is as well,
$\vert {\bf x}({\rm n}) - {\bf x}' \vert < \Delta x$.
Given that we impose the detailed balance condition, then, these
reflecting boundary conditions provide a valid Monte Carlo move from an old
point in the allowed simplex to a new point in the
allowed simplex.

These reflecting boundary conditions can be interpreted in a 
geometrical way.  Essentially, the reflections perform a
type of billiards in the $d-1$ dimensional simplex.
  In each attempted move, we chose a displacement
${\bf x} - {\bf x}({\rm o})$.   We can interpret the
composition variable as a small ball that
moves at a constant speed along this trajectory.
The motion is continued until either a boundary is encountered or
the entire length of the trajectory has been traveled.  If a boundary
of the simplex is encountered, the ball reflects off the hyper-plane
by Newtonian mechanics,  and the trajectory
continues along the new, reflected direction.  The motion, including
possible additional reflections, is continued
until the ball has traveled a distance equal to
the chosen length of the displacement,
$\vert {\bf x} - {\bf x}({\rm o}) \vert$.
The location of the ball at the end of the trajectory
then gives the new point ${\bf x}({\rm n})$.
The reverse move is given by exactly the reverse motion, which
has the same total displacement.

\bibliography{zeolitecchem}
\newpage

\end{document}